\documentclass[12pt]{JHEP3}
\let\ifpdf\relax
\usepackage{amsmath,amsfonts,amssymb,amsthm,amstext,bbm, amscd,eucal}
\usepackage{epsfig}
\usepackage{graphicx}
\usepackage{psfrag}
\usepackage{epstopdf}
\newcommand{\be}{\begin{equation}}
\newcommand{\ee}{\end{equation}}
\newcommand{\bea}{\begin{eqnarray}}
\newcommand{\eea}{\end{eqnarray}}
\newcommand{\bse}{\begin{subequations}}
\newcommand{\ese}{\end{subequations}}
\newcommand{\bi}{\begin{itemize}}
\newcommand{\ei}{\end{itemize}}

\newcommand\vp{\varphi}
\newcommand\hchi{\hat{\chi}}
\newcommand\leff{\lambda_{\rm eff}}

\newcommand{\mpl}{M_{\rm pl}}

\title{\normalsize{\Large{Gravitational Waves from Preheating in M-flation}}}

\author{Amjad Ashoorioon \footnote{a.ashoorioon@lancaster.ac.uk}\\
{Physics Department, Lancaster University, Lancaster, LA1 4YB, United
Kingdom}}

\author{Brandon Fung \footnote{b6fung@uwaterloo.ca}\\ {Department of Physics and Astronomy, University of Waterloo, Waterloo,
Ontario, N2L 3G1, Canada}}

\author{Robert B. Mann\footnote{rbmann@uwaterloo.ca}  \\
{Department of Physics and Astronomy, University of Waterloo, Waterloo,
Ontario, N2L 3G1, Canada} and
{Perimeter Institute for Theoretical Physics, Waterloo, Ontario, N2L
2Y5, Canada}}

\author{Marius Oltean\footnote{moltean@physics.mcgill.ca}\\ {Department of Physics, McGill University, Montr\'{e}al, Qu\'{e}bec, H3A 2T8,
Canada}}

\author{ M. M. Sheikh-Jabbari  \footnote{jabbari@theory.ipm.ac.ir}\\
 School of Physics, Institute for Research in Fundamental Sciences (IPM), Tehran, Iran}

\abstract{Matrix inflation, or M-flation, is a string theory motivated inflationary
model with three scalar field matrices and gauge fields  in the adjoint representation
of the $\mathbf{U}(N)$ gauge group. One of these $3N^2$ scalars appears as the effective
inflaton while the rest of the fields (scalar and gauge fields) can play the role of isocurvature fields during inflation and preheat fields afterwards. There is a region in parameter space and initial field values, ``the hilltop region,''  where predictions of the model
are quite compatible with the recent \textit{Planck} data.
We show that in this hilltop region, if the inflaton ends up in the supersymmetric vacuum, the model can have an embedded preheating mechanism.
Couplings of the preheat modes are related to the inflaton self-couplings and therefore are known from the CMB data. Through lattice simulations performed
using a symplectic integrator, we numerically compute the power spectra
of gravitational waves produced during the preheating stage following
M-flation.  The preliminary numerical simulation of the spectrum from multi-preheat fields peaks in the GHz band with  an amplitude $\Omega_{\mathrm{gw}}h^{2} \propto 10^{-16}$, suggesting that the model has concrete predictions
for the ultra-high frequency gravity-wave probes. This signature could be used to distinguish the model from rival inflationary models.
}

\preprint{\today }

\begin{document}

\parskip 6 pt
\lineskip 2pt

\section{Introduction}

Cosmological observations---evidenced, most notably, by recent data
\cite{Planck} from the \textit{Planck} satellite---are best explained if we have a period of accelerated expansion, inflation \cite{Inflation-Books}, in the early Universe. Models of inflation usually involve one or more scalar fields coupled to (Einstein) gravity, though it is also possible that inflation is driven by gauge fields \cite{gauge-flation-review}.  These models are specified by the form of their kinetic terms as well as the potential. It is more common to take the canonical kinetic term and define the model by its potential(s), even though inflationary models can be realized with non-canonical kinetic terms \cite{ArmendarizPicon:1999rj}.

To explain the observed Universe, inflation should of course end and the energy stored in the inflationary sector should be transferred into the (beyond) Standard Model (SM) particles, an epoch known as the  reheating era \cite{Inflation-Books,reheat}. Perturbative decay of quantum fluctuations of an inflaton field (perturbative reheating) is usually not sufficiently fast and efficient, and leads to reheat temperatures that are  too low to solve particle physics problems and hence to describe what we see\footnote{Big Bang Nucleosynthesis (BBN) requires a temperature of order $1-10$ MeV and baryogenesis requires a temperature of order $1-10$ TeV. The reheat temperature should be at least bigger than these two temperatures.}. One must therefore equip inflationary models with non-perturbative mechanisms of decay that yield sufficiently high  reheat temperatures.     In this context the inflaton field condensate can provide ``time dependent mass terms'' for other fields coupled to it, the \emph{preheat} fields. This more efficient energy transfer mechanism to other (beyond Standard Model) fields, preheating, happens because of possible resonance bands \cite{preheat}. The energy in the preheat fields  will eventually equilibrate or thermalize through usual (perturbative) scattering processes \cite{reheat}.

Observable effects of inflationary models  (in particular the observed CMB aniso\-tropy \cite{Planck}) are usually attributed to what  happened during inflation and are related to super-horizon quantum fluctuations of inflaton fields that appear as classical background fluctuations long after inflation, and after the preheating and reheating eras \cite{Inflation-Books}. The CMB data have hence been used to restrict inflationary models \cite{Planck}.  However the recent \textit{Planck} mission data -- in particular non-observation of non-Gaussianity -- means that the CMB data does not provide sufficient constraints   to specify the inflaton potential.  Other sources of cosmic data must be sought out.

To this end, a more concrete understanding/modelling of reheating and preheating may be needed.  Various inflationary models
can be constrained  by probing possible specific features they left during preheating or reheating. Previous analysis indicates that preheating may have detectable traces on CMB only for a specific class of exotic models  \cite{Bassett}.
If preheating occurs the turbulent, explosive and non-thermal energy transfer to the preheat sector can in principle have possible observable effects by producing a stochastic background of gravity waves typically in $10^7-10^9$ Hz frequency band \footnote{Preheating can also lead to production of long-lived non-linear excitations of the scalar field which dominates the universe and can lead to stochastic gravitational wave background \cite{Zhou:2013tsa}.} \cite{Khlebnikov:1997di}.

The simplest scalar-driven slow-roll  models (in particular, those
with concave potentials\footnote{Note that a choice of non-Bunch-Davies (excited) initial states for the cosmic perturbations can readily change this conclusion \cite{Initial-condition}. Also if gravity is an inherently a classical theory, there will be no B-mode polarization in the CMB \cite{Ashoorioon:2012kh}.}) have so far passed the test very well insofar
as \textit{Planck} results are concerned, see e.g. \cite{After-Planck}. Nevertheless there remain with these models
a plethora of unresolved theoretical difficulties  \cite{Brandenberger}. For instance, to have a successful slow-roll inflation
we need to keep the inflaton mass hierarchically smaller than the Hubble scale $H$ during inflation and quantum corrections to the inflaton potential should not spoil its flatness \cite{Ashoorioon:2011aa}. Moreover, in the class of large-field models there is
also the problem of super-Planckian field excursions: that inflaton(s) in these scalar models typically have field displacements
(in the last $60$ e-folds) many times larger than $\mpl$, in which case quantum (gravity) effects
may become important \cite{Lyth:1996im}.

It is a general belief that these and other theoretical issues regarding possible classical or quantum instabilities in an inflationary model
can/should be addressed within a quantum gravity setup that is operative at some high energy (Planckian or sub-Planckian) scale. Despite providing a richer framework for inflationary model building and for addressing the above mentioned issues, being farther from SM physics, it becomes more challenging in the quantum gravity setups to make connections with physics after inflation and in particular to have a successful reheating scenario. Nonetheless, working within a string theoretic perspective, besides providing a framework to address questions about UV stability and completeness of inflationary models, usually brings another feature: there are many more fields besides the inflaton in the model. These fields can appear as isocurvature entropy modes, affecting the CMB directly, or can appear as preheat fields,  affecting the production of primordial gravity waves in large frequency bands.

M-flation, which we will consider in this work, is one such model \cite{M-flation}.
Although motivated from string theory (quantum gravity) M-flation, as we will show, has the advantage of having an embedded successful preheating mechanism in some regions of parameter space. Furthermore, the model  is based on a gauge field theory, the same framework upon which beyond SM models operate, and is thus close to particle physics setups too.

In general and in a string theory/supergravity framework,  depending on whether the inflaton field(s) is (are) coming from open string or closed string degrees of freedom, there are two venues for inflationary model building \cite{string-models}. M-flation, in this sense, is an open string model. However it has its own specific features that may justify viewing it as a third venue. For example, as we will review in section \ref{sec:M-flation}, inflation in M-flation is not associated with a mobile brane, unlike all the other known open string models. M-flation is
rather motivated by the dynamics of D3-branes subject to a proper RR six-form in a specific ten-dimensional type IIB supergravity background \cite{M-flation}. The inflaton fields of M-flation are three $N\times N$ matrix valued scalar fields associated with the position of a stack of $N$ D3-branes in this background.  The action for M-flation, \emph{cf.} section 2, will hence include
$U(N)$ gauge fields (and possibly their spinorial counterparts in a supersymmetric setting). The model is compatible with the \textit{Planck} data if inflation happens in the hilltop  $\phi<\mu$ region. In the symmetry breaking region, $\phi>\mu$, the model predicts a large tensor/scalar ratio, $r\simeq 0.2$, which is not compatible with the upper bound of $0.11$ with $95\%$ CL if one assumes that the perturbations start from a Bunch-Davies vacuum. The model could be still made compatible with \textit{Planck} if we assume  excited initial states for the scalar or tensor fluctuations, as pointed out in \cite{Initial-condition}.

What renders M-flation  theoretically appealing  is not only its ability to
naturally address and resolve  the theoretical difficulties of standard inflationary
scenarios raised above \cite{gauged-M-flation}, but also the fact
that it can connect to post-inflation physics:
it comes with its own built-in preheating mechanism in some regions of parameter space with no extra parameters (compared to the inflationary background sector), and also it has the desirable form of a gauge theory (\emph{cf.} discussions above).

While work on M-flation  has so far been  directed more toward
exploring it during inflation   \cite{M-flation,gauged-M-flation,M-flation-Landscape}, it is of appreciable importance
to also address the question of its possible observable effects coming from its built-in preheating period.
In particular, we focus our attention in this paper on gravity
waves (GW) produced during the preheating phase following inflation, in some region of parameter space.
Their observational signature is revealed by way of their power spectrum,
which we numerically compute here with the help of the lattice simulator
HLattice 2.0 \cite{Zhiqi}.

The rest of this paper is organized as follows. In section \ref{sec:M-flation},
we review the basic setup of M-flation. In sections \ref{sec:Preheating} and \ref{sec:Parametric Resonance},
we describe its embedded preheating mechanism. Then, in section \ref{sec:Gravitational-Waves-from},
we proceed with computing the power spectra of GW thereby generated.
Finally, section \ref{sec:Conclusion} presents some concluding remarks.

\section{M-flation}{\label{sec:M-flation}}

Our setting is a 10-dimensional type IIB supergravity background,\footnote{For a detailed specification, the reader is referred to section 8
of \cite{M-flation}.} which is probed by a stack of $N$ D3-branes endowed with Yang-Mills
gauge fields. Thus, there exist 6 spatial dimensions perpendicular
to the D3-branes, whose positions  within this subspace
are represented by 6 $N\times N$ matrices. The role of the inflaton,
according to the original M-flation setup, is assumed by 3 out of 6 matrix degrees
of freedom,\footnote{We assumed the 6 extra-dimensions are compactified on a $CY_3$ or $T^6$ manifold that has two three-cycles, one considerably larger than the other. In principle we can use all $6$ extra dimensions and work with 6 matrices, which could be related  generators of  $SO(6)$ or a subgroup of it.} which we henceforth denote as $\boldsymbol{\Phi}_{i},  {i=1,2,3}$.
The inflaton matrices are, by construction, in the adjoint representation
of the $\mathbf{U}(N)$ gauge group; therefore they are
non-commutative as well as Hermitian.

In principle, the dynamics of these matrices is very
complicated (increasingly so with larger $N$), as one
has any number of possible configurations of the D3-branes within
the chosen background. However, there is a way to simplify the situation
and make it computationally tractable. As we will elaborate in the
next subsection, the classical dynamics of this model can be consistently
truncated to a solution where the $N$ D3-branes are uniformly distributed
along the surface of a 2-sphere (within the 6-dimensional orthogonal subspace),
and their positions on this sphere do not change during
inflation. What instead changes is the sphere's radius, which
thereby plays the role of an effective scalar inflaton.

Aside from the above, many other solutions---that make use of more
of the available (classical) degrees of freedom---are of course possible.
This possibility was considered in \cite{M-flation-Landscape} and generically appears as a
multi field inflationary model. In this work, however, we focus on the single field model where the other
``unused'' degrees of freedom in this particular solution will be identified
with preheat fields after inflation ends.

\subsection{Action and equations of motion}

We work in the $\left(-,+,+,+\right)$ metric signature, and use boldface
to denote matrices of dimension $N$.
The effective $\left(3+1\right)$-dimensional action of M-flation
\cite{gauged-M-flation} comprises Einstein gravity, minimally coupled
to a Yang-Mills gauge field $\mathbf{A}_{\mu}$ and the three inflaton
matrices $\mathbf{\Phi}_{i}$,
\begin{equation}
S=\int\mathrm{d}^{4}x\sqrt{-g}\left\{ \frac{\mpl^{2}}{2}R-\frac{1}{4}\mathrm{Tr}\left(\mathbf{F}_{\mu\nu}\mathbf{F}^{\mu\nu}\right)-
\frac{1}{2}\mathrm{Tr}\left(D_{\mu}\boldsymbol{\Phi}_{i}D^{\mu}\boldsymbol{\Phi}_{i}\right)-V
\left(\boldsymbol{\Phi}_{i},\left[\boldsymbol{\Phi}_{j},\boldsymbol{\Phi}_{k}\right]\right)\right\} ,\label{eq:Action}
\end{equation}
where, as usual, $\mpl=1/\sqrt{8\pi G}$ is the reduced Planck mass,
$\mathbf{F}_{\mu\nu}=2\partial_{[\mu}\mathbf{A}_{\nu]}+\mathrm{i}g_{_{\mathrm{YM}}}[\mathbf{A}_{\mu},\mathbf{A}_{\nu}]$
is the gauge field strength, and $D_{\mu}=\partial_{\mu}+\mathrm{i}g_{_{\mathrm{YM}}}[\mathbf{A}_{\mu},\cdot]$
is the gauge covariant derivative. Moreover, the potential is given
by
\begin{equation}
V\left(\boldsymbol{\Phi}_{i},\left[\boldsymbol{\Phi}_{i},\boldsymbol{\Phi}_{j}\right]\right)=
\mathrm{Tr}\left(-\frac{\lambda}{4}[\boldsymbol{\Phi}_{i},\boldsymbol{\Phi}_{j}][\boldsymbol{\Phi}_{i},\boldsymbol{\Phi}_{j}]+\frac{\mathrm{i}\kappa}{3}\epsilon_{jkl}[\boldsymbol{\Phi}_{k},\boldsymbol{\Phi}_{l}]\boldsymbol{\Phi}_{j}+\frac{m^{2}}{2}\boldsymbol{\Phi}_{i}\boldsymbol{\Phi}_{i}\right),\label{eq:Potential}
\end{equation}
where in \eqref{eq:Action} and \eqref{eq:Potential} there is a sum on repeated $i,j,k$ indices and
the three coupling constants have various stringy meanings: $\lambda=8\pi g_{s}=2g_{_{\mathrm{YM}}}^{2}$
is related to the string coupling $g_{s}$, $\kappa=\hat{\kappa}g_{s}\sqrt{8\pi g_{s}}$
is related to the Ramond-Ramond antisymmetric form strength $\hat{\kappa}$,
and $m$ is a parameter that multiples the three spatial coordinates
along the D3-branes in the metric of the background SUGRA theory \cite{M-flation}.
To ensure a constant dilaton therein, we must also impose the constraint
$\lambda m^{2}=4\kappa^{2}/9$ \cite{M-flation}.

The equations of motion for the scalar and gauge fields that follow
from the action (\ref{eq:Action}) are
\begin{gather}
D_{\mu}D^{\mu}\boldsymbol{\Phi}_{i}+\lambda[\boldsymbol{\Phi}_{j},[\boldsymbol{\Phi}_{i},\boldsymbol{\Phi}_{j}]]-\mathrm{i}\kappa\epsilon_{ijk}[\boldsymbol{\Phi}_{j},\boldsymbol{\Phi}_{k}]-m^{2}\boldsymbol{\Phi}_{i}=0,\label{eq:ScalarEOM}\\
D_{\mu}\mathbf{F}^{\mu\nu}-\mathrm{i}g_{_{\mathrm{YM}}}[\boldsymbol{\Phi}_{i},D^{\nu}\boldsymbol{\Phi}_{i}]=0.\label{eq:GaugeEOM}
\end{gather}

\subsection{Truncation to the $\mathbf{SU}(2)$ sector}

The dynamics determined by the equations of motion (\ref{eq:ScalarEOM})
and (\ref{eq:GaugeEOM}) can generically be quite complicated, but
this may be simplified considerably as follows. Let $ \mathbf{J}_{i}\,, {i=1,2,3}$
denote the three $N\times N$ generators of the $\mathbf{SU}(2)$
algebra, so that $\left[\mathbf{J}_{i},\mathbf{J}_{j}\right]=\mathrm{i}\epsilon_{ijk}\mathbf{J}_{k}$.
Now, we decompose the inflaton matrices into two parts,
\begin{equation}
\boldsymbol{\Phi}_{i}=\hat{\phi}\mathbf{J}_{i}+\boldsymbol{\Psi}_{i},\label{eq:PhiDecomposition}
\end{equation}
one parallel and one perpendicular to the $N\times N$ representation of $\mathbf{SU}(2)$,
respectively (that is $\mathrm{Tr}(\mathbf{J}_i \mathbf{\Psi_i})=0$). It was shown in \cite{M-flation} that if $\boldsymbol{\Psi}_{i}=\dot{\boldsymbol{\Psi}}_{i}=0$
initially, then (\ref{eq:ScalarEOM}) implies that $\boldsymbol{\Psi}_{i}$
will remain vanishing for all time. Analogously, if $\mathbf{A}_{\mu}$
is also  initially turned off, then the commutator in (\ref{eq:GaugeEOM})
will not source $\mathbf{F}_{\mu\nu}$, and therefore the gauge field always
stays  turned off as well.

Hence, it is possible to consistently restrict the classical dynamics
of this model to a sector where $\boldsymbol{\Psi}_{i}=\mathbf{A}_{\mu}=0$,
so that the inflationary trajectory is determined solely by $\hat{\phi}$,
the length of the inflaton matrices along the direction of $\mathbf{SU}(2)$.
This realizes precisely the picture described earlier of the D3-branes
fixed upon the surface of a 2-sphere with variable radius, now identified
with the value of effective inflaton field  $\hat{\phi}$.

Concordantly, the vanishing $\boldsymbol{\Psi}_{i}$ and $\mathbf{A}_{\mu}$
fields are referred to as \emph{spectators}. Upon setting them to zero, the
action (\ref{eq:Action}) simplifies propitiously to
\begin{equation}
S=\int\mathrm{d}^{4}x\sqrt{-g}\left\{ \frac{\mpl^{2}}{2}R+\mathrm{Tr}\mathbf{J}_{i}^{2}\left(-\frac{1}{2}\partial_{\mu}\hat{\phi}\partial^{\mu}\hat{\phi}-\frac{\lambda}{2}\hat{\phi}^{4}+\frac{2\kappa}{3}\hat{\phi}^{3}-\frac{m^{2}}{2}\hat{\phi}^{2}\right)\right\} ,\label{eq:PhiHatAction}
\end{equation}
where $\mathrm{Tr}\mathbf{J}_{i}^{2}=N(N^{2}-1)/4$, using the properties
of $\mathbf{SU}(2)$. Performing a field redefinition $\phi=\sqrt{\mathrm{Tr}\mathbf{J}_{i}^{2}}\hat{\phi}$
brings the inflaton to a canonically normalized form, yielding
\begin{equation}
S=\int\mathrm{d}^{4}x\sqrt{-g}\left\{ \frac{\mpl^{2}}{2}R-\frac{1}{2}\partial_{\mu}\phi\partial^{\mu}\phi-V_{0}\left(\phi\right)\right\} \label{eq:PhiAction}
\end{equation}
which is  the familiar single scalar field inflationary action.

Defining effective couplings $\lambda_{\mathrm{eff}}\equiv8\lambda/N(N^{2}-1)$
and $\kappa_{\mathrm{eff}} \equiv2\kappa/\sqrt{N(N^{2}-1)}$ and then using
the constraint that the background is a solution to the supergravity equations of motion with constant dilaton, $\lambda m^{2}=4\kappa^{2}/9$, the
effective potential can be written, with $\mu\equiv \sqrt{2}m/\sqrt{\lambda_{\mathrm{eff}}}$,
simply as
\begin{equation}
V_{0}\left(\phi\right)=\frac{\lambda_{\mathrm{eff}}}{4}\phi^{4}-\frac{2\kappa_{\mathrm{eff}}}{3}\phi^{3}+\frac{m^{2}}{2}\phi^{2}=\frac{\lambda_{\mathrm{eff}}}{4}\phi^{2}\left(\phi-\mu\right)^{2}.\label{eq:EffectivePotential}
\end{equation}

Thus, in the $\mathbf{SU}(2)$ sector, the inflationary
potential of M-flation assumes the form of a symmetry-breaking potential.
It has two global minima: one at $\phi=\mu$ (corresponding to a supersymmetric
vacuum, when the $N$ D3-branes blow up into a giant D5-brane wrapping a fuzzy two sphere) and
one at $\phi=0$ (corresponding to the trivial solution, when the
matrices become commutative). For typical inflationary trajectories
determined by this potential, all necessary parameters can be obtained
by demanding certain standard requirements (namely, 60 e-foldings
of inflation, together with a COBE normalization of $\delta_{H}\simeq2.41\times10^{-5}$
and a spectral index of $n_{s}=0.96$). The resultant numerical values are as follows. Further details about this analysis and the corresponding slow-roll trajectories in M-flation may be found in \cite{M-flation,M-flation-Landscape};  here we just quote the results.

{\bf (a)}~$ \phi_{i}> \mu$

Suppose inflation starts when $ \phi_{ i}> \mu$.  The aforementioned standard requirements imply
 \be \label{case1}%
 \phi_{i} \simeq 43.57
\mpl \,, \qquad \phi_{f} \simeq 27.07 \mpl \,,\quad  \quad \mu
\simeq 26 \mpl\,.%
 \ee%
  and
\begin{equation}\label{phi>mu}
   \lambda_{\rm eff}\simeq 4.91 \times 10^{-14}, \quad m\simeq 4.07\times 10^{-6} \mpl, \quad \kappa_{\rm eff}\simeq 9.57 \times 10^{-13} \mpl.
\end{equation}

Taking $n_S\simeq 0.96$, the tensor/scalar ratio turns out to be $0.2$ which is outside the $2\sigma$ allowed region of \textit{Planck } in the $n_S-r$ plane. One  can render this region of M-flationary phase space compatible with the data by assuming the modes start from a non-Bunch-Davies vacuum \cite{Initial-condition}.

{\bf (b)}~$ \mu/2<\phi_{i}< \mu$

To fit the observational constraints we find
\be \label{casetwo}%
\phi_{i} \simeq 23.5 \mpl \,,  \qquad \phi_{f} \simeq 35.03 \mpl \,,
\qquad \mu \simeq 36 M_{P}\,.%
\ee%
and%
\begin{equation}\label{phi<mu}
\lambda_{\rm eff}\simeq 7.18\times 10^{-14}\,, \qquad m\simeq 6.82\times
10^{-6} \mpl\,, \qquad \kappa_{\rm eff} \simeq 1.94\times 10^{-12}
\mpl\,.
\end{equation}%

{\bf (c)}~$ 0<\phi_{i}< \mu/2$

In this case we obtain
\be \label{case2}%
\phi_{i} \simeq 12.5 \mpl \,,  \qquad \phi_{f} \simeq 0.97 \mpl \,,
\qquad \mu \simeq 36 M_{P}\,.%
\ee%
and%
\begin{equation}\label{phi<muover2}
\lambda_{\rm eff}\simeq 7.18\times 10^{-14}\,, \qquad m\simeq 6.82\times
10^{-6} \mpl\,, \qquad \kappa_{\rm eff} \simeq 1.94\times 10^{-12}
\mpl\,.
\end{equation}%

Due to the $\phi\to \mu-\phi$ symmetry of the background, the curvature perturbations in regions ${\bf (b)}$ and ${\bf (c)}$ turn out to have the same spectral tilt  $n_S=0.96$ and tensor-to-scalar ratio $r=0.048$. These predictions are within the $1\sigma$ region of \textit{Planck}-allowed parameter space.  The two regions ${\bf (b)}$ and ${\bf (c)}$, however,  could be distinguished by their predictions for the amplitude of isocurvature perturbations at the Hubble scale \cite{M-flation}, as the masses of isocurvature modes do not satisfy the symmetry $\phi\rightarrow \mu-\phi$, which the classical background enjoys. As we will see, in region ${\bf (c)}$ the model has an embedded preheating mechanism that leads to observable gravity waves in the high frequency region.

In this way, M-flation resolves all of the problems raised earlier
vis-\`{a}-vis single scalar field inflation for several reasons. First, its effective couplings
can easily be made naturally small, provided $N$ is chosen to be
sufficiently large. For example, $N\approx48\,000$ D3-branes turn
out to suffice in this case for ameliorating any hierarchy problem.
Second, the total amount of field displacement during M-flation
has been argued \cite{gauged-M-flation} to be less than the UV cutoff
of this model, so there is no trans-Planckian problem.  Finally, this approach suggests a clear physical meaning for the inflaton, namely the radius of the two-sphere on which D3-branes live.

Despite its theoretical successes, M-flation has not been up to now
extensively exploited in terms of deriving observationally testable
predictions that may help set it aside from rival inflationary models.
This is what we turn our attention to next, in the context of preheating.

\section{Preheating in M-flation}\label{sec:Preheating}

The preheating mechanism after inflation in typical models of inflation necessitates the
introduction of one or more extra matter fields, or preheat fields, into which the inflaton presumably
ought to decay \cite{preheat}. M-flation comes with this feature tacitly built-in,
by way of its spectators $\boldsymbol{\Psi}_{i}$ and $\mathbf{A}_{\mu}$.
Although, as discussed, these are assumed to be turned off classically,
they can nevertheless be excited quantum mechanically. During inflation, these quantum fluctuations can cross the horizon and can become observable as isocurvature perturbations. The amplitude of the largest modes in each inflationary region was computed in \cite{M-flation}, and shown to be generically too small to have observable effects. After inflation, however, they appear as preheat fields which can have observable effects on the GWs produced in this era.

To this end, we need to study the equations of motion and quantize their solution. We hence start  with  $\hat{\boldsymbol{\Psi}}_{i}$ and
$\hat{\mathbf{A}}_{\mu}$ as perturbations in the action (\ref{eq:Action})---with the hats denoting ``quantumness''--- and deduce
the resulting equations of motion. As usual in inflationary cosmic perturbation theory we assume these perturbations to be of the same order and both be much smaller than the background field values and hence keep only the first order terms in these perturbations in the equations of motion. In either case, these will take
the expected form of Mathieu equations suitable for preheat fields.
We discuss each case separately.

\subsection{Scalar preheat fields}

Setting $\hat{\mathbf{A}}_{\mu}=0$ and expanding (\ref{eq:Action})
to quadratic order in $\hat{\boldsymbol{\Psi}}_{i}$, we get \cite{gauged-M-flation}:
\begin{gather}
S_{\boldsymbol{\Psi}}^{(2)}=\int\mathrm{d}^{4}x\sqrt{-g}\left\{ -\frac{1}{2}\mathrm{Tr}\left(\partial_{\mu}\hat{\boldsymbol{\Psi}}_{i}\partial^{\mu}\hat{\boldsymbol{\Psi}}_{i}\right)-\frac{1}{2}M_{\boldsymbol{\Psi}}^{2}\left(\phi\right)\mathrm{Tr}\left(\hat{\boldsymbol{\Psi}}_{i}^{2}\right)\right\} ,\label{eq:QuadraticActionPsi}
\end{gather}
where there are two solutions for the scalar spectator masses, dubbed
$\alpha$-modes and $\beta$-modes respectively:
\begin{equation}
M_{\boldsymbol{\Psi}}^{2}\left(\phi\right)=\begin{cases}
M_{\alpha_{j}}^{2}\left(\phi\right)=\frac{1}{2}\lambda_{\mathrm{eff}}\phi^{2}(j+2)(j+3)-2\kappa_{\mathrm{eff}}\phi(j+2)+m^{2}, & 0\leq j\leq N-2,\\ \,\,\ \\
M_{\beta_{j}}^{2}\left(\phi\right)=\frac{1}{2}\lambda_{\mathrm{eff}}\phi^{2}(j-1)(j-2)+2\kappa_{\mathrm{eff}}\phi(j-1)+m^{2}, & 1\leq j\leq N,
\end{cases}\label{eq:MassSpectrumScalar}
\end{equation}
with degeneracy $2j+1$ for each mode. The above $\phi-$dependent masses, besides a bare mass, induce both types of $\phi^2 \chi^2$ and $\phi\chi^2$ interactions for the preheat fields $\chi$.

It can be easily shown that if inflation happens in the region (\textbf{c}), the above masses for $\alpha-$ and $\beta-$ modes become tachyonic for an interval during the preheating era, if  $j>j_{\rm min}$. For $\alpha$-modes, $j_{\rm min}= 94$ and for $\beta-$ modes $j_{\rm min}=16$. For these modes, we have to alleviate the problem by including the corrections up to quartic order in $\hat{\boldsymbol{\Psi}}_{i}$ \footnote{ The reason for inclusion of these  higher order terms is to stabilize the potential for large $\hat{\boldsymbol{\Psi}}_{i}$; otherwise later simulations for the gravitational waves become unstable. We have neglected the cross-coupling that may arise from the interactions of the gauge mode and spectator mode at lower order.}. We get:
\begin{equation}
\begin{split}
S_{\boldsymbol{\Psi}}^{(3)}&=\int\mathrm{d}^{4}x\sqrt{-g}\left\{ -\mathrm{K}_{\boldsymbol{\Psi}}\left(\phi\right)\mathrm{Tr}\left(\hat{\boldsymbol{\Psi}}_{i}^{3}\right)\right\}\, ,\\ S_{\boldsymbol{\Psi}}^{(4)}&=\int\mathrm{d}^{4}x\sqrt{-g}\left\{ -\Lambda_{\boldsymbol{\Psi}}\mathrm{Tr}\left(\hat{\boldsymbol{\Psi}}_{i}^{4}\right)\right\} ,\label{eq:CubicQuarticActionPsi}
\end{split}
\end{equation}
with
\begin{equation}
\mathrm{K}_{\boldsymbol{\Psi}}\left(\phi\right)=\begin{cases}
\mathrm{K}_{\alpha_{j-2}}\left(\phi\right)={\displaystyle
\Big[\frac{\kappa_{\mathrm{eff}}}{6}-\frac{\lambda_{\mathrm{eff}}}{4}j\phi\Big] \sqrt{j+1}\ \mathbb{G}_j},
\quad   & 3\leq j\leq N,\\
\,\\
\mathrm{K}_{\beta_{j+2}}\left(\phi\right)={\displaystyle
\Big[\frac{\kappa_{\mathrm{eff}}}{6}+\frac{\lambda_{\mathrm{eff}}}{4}\Big(\frac{j+1}{2}\Big)\phi\Big]\sqrt{j}\ \mathbb{G}_j},
\quad  &-1\leq j\leq N-2,
\end{cases}\label{eq:CubicCouplings}
\end{equation}
and\\
\begin{equation}
\hspace*{-1.5cm}\Lambda_{\boldsymbol{\Psi}}=\begin{cases}
\Lambda_{\alpha_{j-2}}={\displaystyle \left(j+1\right)\ \mathbb{U}_j},\  & \quad 3\leq j\leq N,\\
\,\cr
\Lambda_{\beta_{j+2}}={\displaystyle j\ \mathbb{U}_j},\  & \quad -1\leq j\leq N-2\,,
\end{cases}\label{eq:QuarticCouplings}
\end{equation}
where
\be
\begin{split}
\mathbb{G}_j&=12\left(-1\right)^{N+1}\sqrt{N\left(N^{2}-1\right)}\sqrt{j\left(j+1\right)}\ \Big(\begin{array}{ccc}
j & j & j\\
-1 & 0 & 1
\end{array}\Big)\Big\{\begin{array}{ccc}
j & j & j\\
\frac{N-1}{2} & \frac{N-1}{2} & \frac{N-1}{2}
\end{array}\Big\},\\ \,\,\\
\mathbb{U}_j &=\frac{\lambda_{\mathrm{eff}}}{4}N(N^2-1)\ (j+1)\ \sum_{c=0}^{2j}\left(2c+1\right)\Big(\begin{array}{ccc}
j & j & c\\
1 & -1 & 0
\end{array}\Big)^{2}\Big\{\begin{array}{ccc}
j & j & c\\
\tfrac{N-1}{2} & \tfrac{N-1}{2} & \tfrac{N-1}{2}
\end{array}\Big\}^{2}\,,
\end{split}\ee
and $(:::)$ and $\{:::\}$ respectively denote Wigner $3j$ and $6j$ symbols \cite{Wigner-symbol}.

We remark that the cubic couplings (\ref{eq:CubicCouplings}) are
linearly dependent on the inflaton, whereas the quartic ones (\ref{eq:QuarticCouplings})
are manifestly independent   (i.e. they are constants for a
given $j$). Moreover, for reasonable values of $\phi$, it is plain
to see that
\begin{equation}
\frac{\mathrm{K}_{\boldsymbol{\Psi}}}{\mpl}\ll\Lambda_{\boldsymbol{\Psi}},\label{eq:KappaLessLambda}
\end{equation}
in virtue of the fact that the left-hand side is proportional to products
of Wigner symbols, while the right-hand side is proportional to large sums
of products of squares of Wigner symbols\footnote{This claim can be easily checked by explicitly computing the couplings'
numerical values for any given $j$.%
}. Consequently, we can treat the cubic terms as negligible. $\Lambda_{\boldsymbol{\Psi}}$ in general is mode dependent, however, one can show that for large $j$ it becomes $j$-independent and is
\begin{equation}\label{lpsi}
\Lambda_{\boldsymbol{\Psi}}\simeq 1.0069\times 10^{11}\frac{\lambda_{\rm eff}}{4}\,.
\end{equation}
One can therefore take the potential of any scalar ($\alpha$ or $\beta$)
mode $\hat{\chi}$ to be
\begin{equation}
V\left(\phi,\hat{\chi}\right)=V_{0}\left(\phi\right)+\frac{1}{2}M_{\boldsymbol{\Psi}}^{2}\left(\phi\right)\hat{\chi}^{2}+\Lambda_{\boldsymbol{\Psi}}\hat{\chi}^{4}.\label{eq:ScalarPotential}
\end{equation}

Performing the usual Fourier decomposition $$\hat{\chi}\left(t,\mathbf{x}\right)=\int\frac{\mathrm{d}^{3}k}{\left(2\pi\right)^{3/2}}\
\bigl[\chi_{k}(t)\hat{a}_{k}\exp(-\mathrm{i}\mathbf{k}\cdot\mathbf{x})+\chi_{k}^{*}(t)\hat{a}_{k}^{\dagger}\exp(\mathrm{i}\mathbf{k}\cdot\mathbf{x})\bigr]$$
 the corresponding equation of motion can then be written as
\begin{equation}
\ddot{\chi}_{k}+3H\dot{\chi}_{k}+\left(\frac{k^{2}}{a^{2}}+M_{\boldsymbol{\Psi}}^{2}\left(\phi\right)\right)\chi_{k}+
4\Lambda_{\boldsymbol{\Psi}}\chi_{k}^{3}=0.\label{eq:ScalarModesEOM}
\end{equation}
As we will see, this has the familiar form of a Mathieu equation in
the regime where $\phi$ is oscillating about the vacuum (modulo the
last term which, as discussed, was included  to keep
the potential bounded from below), and can therefore lead to parametric
resonance.

\subsection{Gauge preheat fields}

The story here proceeds along similar, albeit slightly simpler lines.
Setting $\hat{\boldsymbol{\Psi}}_{i}=0$ and expanding (\ref{eq:Action})
to quadratic order in $\hat{\mathbf{A}}_{\mu}$ yields \cite{gauged-M-flation}:
\begin{equation}
S_{\mathbf{A}}^{(2)}=\int\mathrm{d}^{4}x\sqrt{-g}\left\{ -\mathrm{Tr}\left(\partial_{[\mu}\hat{\mathbf{A}}_{\nu]}\partial^{[\mu}\hat{\mathbf{A}}^{\nu]}\right)-\frac{1}{2}M_{\mathbf{A}}^{2}\left(\phi\right)\mathrm{Tr}\left(\hat{\mathbf{A}}_{\mu}^{2}\right)\right\} ,\label{eq:QuadraticActionA}
\end{equation}
where the mass spectrum is given by
\begin{equation}
M_{\mathbf{A}}^{2}\left(\phi\right)=\frac{1}{4}\lambda_{\mathrm{eff}}\phi^{2}j(j+1),\qquad0\leq j\leq N-1.\label{eq:MassSpectrumGauge}
\end{equation}
The degeneracy for $j=0$ is 2 (corresponding to massless gauge fields) while for $j\geq 1$ is  $3(2j+1)$, the factor of three corresponding to the three polarizations of a four dimensional massive vector field. Unlike the scalar case,
though, because (\ref{eq:MassSpectrumGauge}) only contains a $\phi^{2}$
term, we need not worry about the danger of acquiring tachyonic masses
and the higher order corrections will always remain small compared the leading quadratic terms.\footnote{Note  that massless gauge field states do not couple to the background effective inflaton (as the effective inflaton is a real field and massless gauge fields are in the center $\mathbf{U}(1)$ of the $\mathbf{U}(N)$ gauge symmetry. The $\mathbf{U}(N)$ gauge symmetry is spontaneously broken to $\mathbf{U}(1)$ by the background field configuration.} We can therefore safely ignore all higher-order corrections and
write the equation of motion for the Fourier modes $A_{k}$ of the
gauge preheat fields as
\begin{equation}
\ddot{A}_{k}+H\dot{A}_{k}+\left(\frac{k^{2}}{a^{2}}+M_{\mathbf{A}}^{2}\left(\phi\right)\right)A_{k}=0.\label{eq:GaugeModesEOM}
\end{equation}
Despite the fact that  the Hubble friction term appears with a different coefficient
than in the scalar case (\ref{eq:ScalarModesEOM}), we still
get a Mathieu equation when the inflaton $\phi$ oscillates around its minimum toward the end of inflation.

The next question to ask is then what the parametric resonance idiosyncratic
to (\ref{eq:ScalarModesEOM}) and (\ref{eq:GaugeModesEOM}) can give
us. A potentially rich and predictive product thereof is GW production.

\section{Parametric resonance \label{sec:Parametric Resonance}}

\subsection{SUSY-breaking vacuum}

If the initial condition  is such that inflation happens in regions ({\bf a}) or ({\bf b}), the inflaton will finally end up oscillating around the SUSY-breaking vacuum, $\phi=\mu$. It might be thought the inflaton oscillations around the  vacuum,  $\phi=\mu$, and its couplings to different preheat fields can create parametric resonance. However, it can be shown that the rest masses of $\alpha$ and $\beta$ modes in this region are so large that non-adiabatic particle production is suppressed. To be specific, let us focus on $\alpha-$modes and $\beta-$modes. A similar analysis and argument could  be repeated for the gauge modes as well.

The mass functions for the $\alpha$ and $\beta$ modes can be unified in the following form
\begin{equation}\label{M-unified}
 M_{\boldsymbol{\Psi}}^{2}\left(\phi\right)=\frac{1}{2}\lambda_{\mathrm{eff}}\omega(\omega-1)\phi^2+
 2\kappa_{\mathrm{eff}}\phi\omega+m^{2},
\end{equation}
where
\begin{equation}\label{GWprofile}
 \omega=\left\{
\begin{array}{ll}
-(j+2)\qquad 1\leq j\leq N-2,  &\\
(j-1)\qquad 1\leq j\leq N. &
\end{array}
\right.
\end{equation}
Expanding the interaction term around the SUSY-breaking vacuum $\phi=\mu$  and introducing the variable $\vp\equiv\phi-\mu$, the interaction term between the inflaton and spectators looks like\footnote{In the rest of the analysis we will drop the quartic $\Lambda_{\mathbf{\Psi}} \hchi^4$ term. As we will see in the next subsection presence of this term weakens the particle production and thus strengthens our results.}
\begin{equation}\label{Vint}
V_{\rm int}=\frac{1}{2} g_{4}^2 \vp^2 \hchi^2+\frac{1}{2} g_3 \vp \hchi^2 +\frac{1}{2} m_{\hchi}^2\hchi^2,
\end{equation}
where
\be\begin{split}
  g_4^2 &= \frac{\lambda_{\rm eff} (\omega^2-\omega)}{2}\,,\\
  g_3 &= \frac{1}{2}\lambda_{\rm eff} \mu (2\omega^2+\omega)\,,\\
 m_{\hchi}^2&= \frac{\lambda_{\rm eff} \mu^2}{2}(\omega+1)^2=m^2 (1+\omega)^2\,,
\end{split}\ee
and $\varphi$ varies between zero and $\Phi=\mu-\phi_f\simeq 1 \mpl$.  Despite the existence of interactions like $\varphi\hchi^2$, since the rest masses of all the $\hchi$ fields are larger or equal to the mass of the inflaton, perturbative decay of the inflaton to none of the $\hchi$ fields is possible.\footnote{For the same reason the tachyonic resonance of \cite{Tachyonic-preheating} does not occur in our case.}

Around the SUSY-breaking vacuum, the inflaton potential to a large extent resembles $\frac{1}{2}m^2 \vp^2$. Therefore, the inflaton has an oscillatory behavior  $\vp(t)\approx \Phi \sin(mt)$ \cite{preheat} around the SUSY-breaking vacuum. It can be shown that the contribution of the $g_4^2 \phi^2 \hchi^2$ interaction is subdominant with respect to the $g_3 \phi\hchi^2$ for all $\omega$'s. The ratio of two interactions is
\begin{equation}\label{R}
R\equiv\frac{g_4^2\vp(t)^2\hchi^2}{g_3\vp(t)\hchi^2}\approx \frac{\omega-1}{2\omega+1}\frac{\Phi}{\mu}\sin(mt).
\end{equation}
For all values   of $\omega>0$ this ratio is less than one\footnote{$\omega=0$ (the $j=1$ $\beta$ mode) does not have any interaction with the inflaton.}, since the ratio $\Phi/\mu\lesssim 0.04$ in both the (\textbf{a}) and (\textbf{b}) regions. Thus we will drop this  quartic interaction term in comparison with the cubic one in the rest of the analysis.

Let us analyze \eqref{eq:ScalarModesEOM} in a non-expanding background where $a=1$. Dropping the contribution of the quartic interaction, for an oscillating inflaton the approximated equation takes the form
\begin{equation}\label{approx-eq}
\ddot{\hchi}_{k}+\left(k^2+m_{\hchi}^2 +\frac{\leff \mu\Phi}{2} \omega (2\omega+1) \sin(mt)\right)\hchi_k=0.
\end{equation}
Introducing the new variable $z\equiv \frac{mt}{2}+\frac{\pi}{4}$ and ${}^{\prime}\equiv \frac{d}{dz}$, the equation takes the form of a Mathieu equation \cite{Mc Lachlan}
\begin{equation}\label{app-eq-z}
\hchi^{\prime\prime}+(A_k-2q \cos(2z))\hchi=0,
\end{equation}
where
\begin{eqnarray}
  A_k &\equiv& \frac{4(k^2+m^2)}{m^2}, \\
  q &\equiv& \frac{\leff \mu \Phi \omega(2\omega+1)}{m^2}=\frac{2\Phi}{\mu}\omega(2\omega+1).
\end{eqnarray}
It is known  \cite{Landau} that  equation \eqref{app-eq-z} has solutions with an exponential instability $\hchi\propto \exp(\mu_k^{(n)} z)$ that represent a burst of particle production. The solutions have resonance bands with the width $\Delta k^{(l)}\simeq q^l$. If $q\ll 1$, what is known as narrow resonance band,  the resonance occurs in bands near $A_k\simeq l^2 $, where $l$ is a nonzero integer.
Hence the widest band is the first instability band. Imposing the condition $q<1$ for the inflationary region (\textbf{a}) where $\mu\simeq 26~\mpl$, only $0\leq \omega\leq 2$ ($1\leq j\leq 3$ $\beta$ modes) lead to narrow resonance. In the region (\textbf{b}), where $\mu<36$, besides the aforementioned modes, $\omega=-3$ ($j=1$ $\alpha$ mode) can also lead to narrow resonance. The factor $\mu_k$, the Floquet index, for the first instability band is given by \cite{preheat}
\begin{equation}\label{muk}
\mu_k=\sqrt{\left(\frac{q}{2}\right)^2-(\frac{2k}{m}-1)^2},
\end{equation}
where the resonance happens for the narrow momentum $k$ range $1-\frac{q}{2}\leq \frac{2k}{m}\leq 1+ \frac{q}{2}$. It obtains its maximum at $\mu_k=q/2$ at $k=m/2$.

In an expanding background the redshift of momentum $k$ from the resonance band can prevent the resonance. As pointed out in \cite{preheat}, the condition for the first band to be effective during expansion is
\begin{equation}\label{eff-resonance}
q^2 m\gtrsim H.
\end{equation}
The inequality is not satisfied for the modes that can undergo  parametric resonance in flat space-time. This is because during preheating $H\simeq 0.1 m$ \cite{preheat} and $\Phi^2/\mu^2\lesssim 1.5\times 10^{-3}$. Thus narrow parametric resonance for these modes cannot lead to preheating.

For larger values of $\omega$, the resonance is broad. However, one can show that the large rest mass of these modes, $m_{\hchi}=m(\omega+1)$, and the smallness of the amplitude of oscillations with respect to the supersymmetry-breaking vacuum $\mu$, shuts off the particle production. To see this, let us note that the time-dependent frequency in the equation of motion for $\hchi$ in an expanding background is given by
\begin{equation}\label{Omega-t}
\Omega=\sqrt{\frac{k^2}{a^2}+m_{\hchi}^2+\frac{\leff\mu\Phi}{2}\omega(2\omega+1)\sin(mt)}.
\end{equation}
The condition for the adiabaticity violation is that
\begin{equation}\label{non-adiabaticity}
    \left|\frac{\dot{\Omega}}{{\Omega^2}}\right|\simeq \frac{1}{2}\frac{\omega(2\omega+1)\cos(mt)}{m ((\omega+1)^2-\omega(2\omega+1)\frac{\Phi}{\mu}\sin(mt))^{3/2}}\frac{\Phi}{\mu}\gtrsim 1,
\end{equation}
a condition that cannot be satisfied for large values of $\omega$ due to the smallness of $\Phi/\mu$. Similar arguments can be given for the gauge spectator modes.

Recapitulating our results, it is not possible to reheat M-flation around the SUSY-breaking minimum via any of the $\alpha$, $\beta$ or gauge spectators modes. The supersymmetric model is equipped with fermionic spectators that might contribute to this process.
Nonetheless, due to Pauli exclusion, resonances cannot happen for fermionic modes and considering them will not change the above result.

\subsection{Supersymmetric vacuum\label{sec:SUSY-vac}}

Unlike the supersymmetry breaking vacuum, parametric resonance around $\phi=0$ (supersymmetric vacuum) can be quite effective through the spectator modes. We first focus on the scalar preheat fields. The equation of motion for the perturbations $\mathbf{\Psi}_i$ can be decomposed into the equation of motion for the $\alpha$ and $\beta$ spectator modes which in Fourier space takes the form
\begin{equation}\label{eom-scalar-spectators}
\ddot{\hchi}_k+3H\dot{\hchi}_k+\left(\frac{k^2}{a^2}+\frac{\leff}{2}\phi^2 (\omega^2-\omega)+\frac{3}{2}\mu\lambda\omega\phi+m^2\right)\hchi_k+4\Lambda_{\boldsymbol{\Psi}}\hat{\chi}_k^{3}=0.
\end{equation}
The bare masses of the spectator modes are equal to the inflaton mass $m^2$ and in principle for large values of $\omega$, the adiabatic condition  may be broken violently. However, as we will see, self-interactions of the $\hchi$ particles, incorporated in the last term of the equation of motion, slows down the parametric resonance.

In terms of the dimensionless time variable $\tilde z$, defined as
\begin{equation}\label{tprime}
\tilde{z}\equiv mt,
\end{equation}
the equations of motion for the inflaton and the background are
\begin{eqnarray}\label{eom-infl}
\phi^{\prime\prime}+3 \mathcal{H}\phi^{\prime}+\left(\frac{2\phi^3}{\mu^2}-\frac{3\phi^2}{\mu}+\phi\right)=0,\\
\mathcal{H}^2=\frac{1}{3\mpl^2}\left[\frac{1}{2}{\phi^{\prime}}^2+\frac{1}{2}\phi^2 \left(\frac{\phi}{\mu}-1\right)^2\right],
\end{eqnarray}
where
\begin{equation}\label{Hscript}
\mathcal{H}\equiv \frac{a^{\prime}}{a}.
\end{equation}
The equation of motion for the Fourier mode, $\mathcal{X}_k\equiv a^{3/2} \hchi_{k}$, is
\begin{equation}\label{X-eom}
\mathcal{X}_k^{\prime\prime}+{\Omega_{k}}^2 \mathcal{X}_k+\frac{4\Lambda_{\boldsymbol{\Psi}}}{a^3 m^2} \mathcal{X}_k^3=0,
\end{equation}
where
\begin{equation}\label{omegal}
{\Omega_{k}}^2\equiv \frac{k^2}{m^2 a^2}+\frac{\phi^2}{\mu^2}(\omega^2-\omega)+\frac{3\phi}{\mu}\omega+1-\frac{3}{4}{\mathcal{H}}^2-\frac{3}{2}\frac{a^{\prime\prime}}{a}.
\end{equation}
Eq.\eqref{X-eom} can be solved imposing the Bunch-Davies vacuum on the mode $\mathcal{X}_k$
\begin{equation}\label{BD-vac}
\mathcal{X}_k\rightarrow \frac{e^{-i \frac{\Omega_{k} t^{\prime}}{m}}}{\sqrt{2\Omega_{k}}}
\end{equation}
at the beginning of   preheating.  The
number density for the produced particles is \cite{preheat}
\begin{equation}\label{n-ell}
n^{\mathcal{X}}_{k}=\frac{\Omega_{k}}{2}\left(m^2 \frac{|\mathcal{X}_k^{\prime}|^2}{\Omega_{k}^2}+|\mathcal{X}_k|^2\right)-\frac{1}{2}.
\end{equation}

\begin{figure}[t]
\includegraphics[angle=0,
scale=0.80]{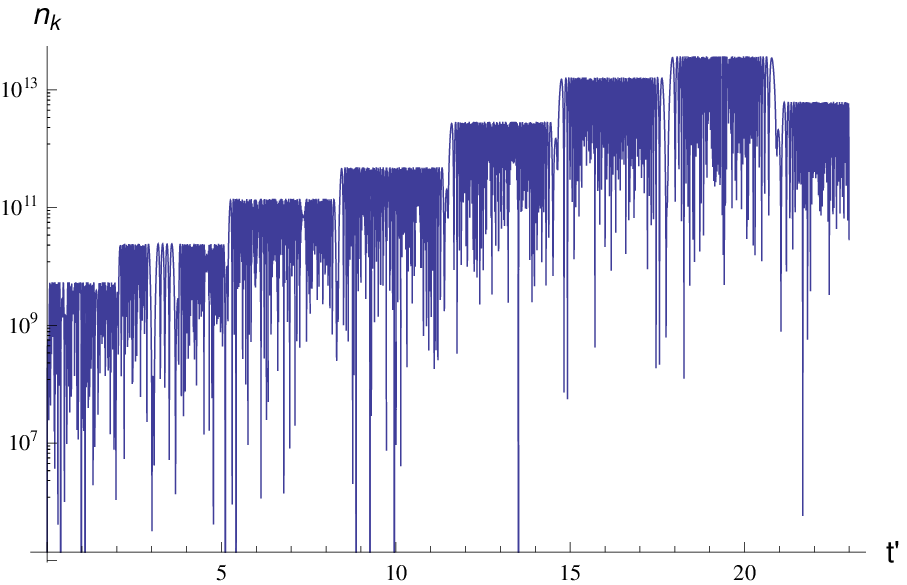}
\includegraphics[angle=0,
scale=0.80]{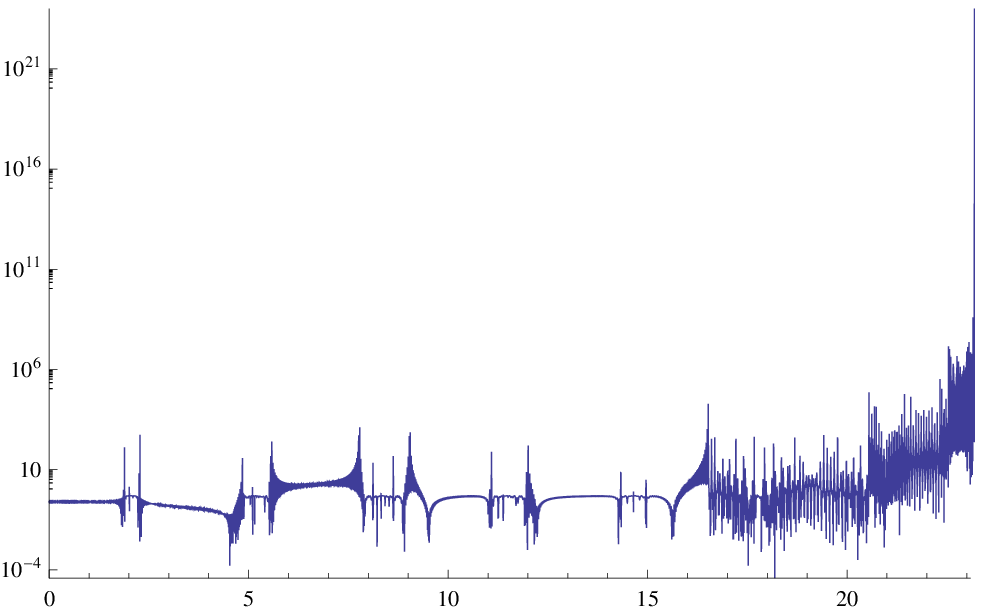}
\caption{Left graph shows how the comoving number density of the $\hchi$ particles, $n^{\mathcal{\chi}}_k$ evolves as a function of $\tilde{z}$ for $k=0$, in the absence of the quartic self-coupling term, which explicitly exhibits the stochastic resonance behavior. The right figure shows the same when the quartic coupling term is added to the Lagrangian of the $\hchi$ field. As can be seen, the self-coupling term slows down the resonance.}
\label{fig-resonance-scalar}
\end{figure}
To demonstrate the contribution of the cubic term to the comoving number density,  we have numerically solved the equations for perturbation in the presence and absence of the cubic contribution to the equations of motion (\ref{X-eom}) for  $k=0$ for the largest $j$ $\beta-$mode. As can be seen in the L.H.S. graph of Fig. \ref{fig-resonance-scalar}, in the absence of the cubic term, the number density of the produced particles exhibits  stochastic resonance behavior \cite{preheat}, {\it i.e.} it typically increases at some specific moments but it may decrease as well. In between these instants, the number density remains approximately constant (sharp oscillations on the plateaus are only numerical artifacts). The interval between the kicks in $n_k$ is roughly about $\pi$, which is the small interval in which the mode becomes massless and tachyonic. However, once the cubic term (from the quartic self-coupling term)
is added to the equation of motion (\ref{X-eom}), $n_k$ ceases to exhibit resonance behaviour initially, its value being highly suppressed. This continues until the cubic term in the equation of motion of the scalar spectator redshifts and the mode revert to  resonance behaviour.
\begin{figure}[t]
\begin{center}
\includegraphics[angle=0,
scale=1]{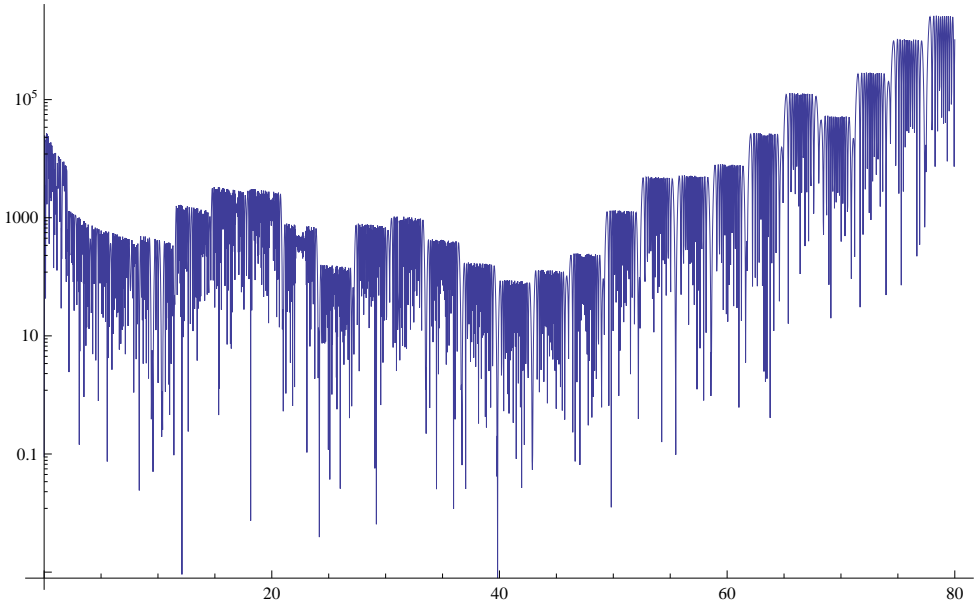}
\caption{$n_k^{\mathcal{A}}$ vs. $\tilde{z}$. Despite the decrease in the number density of the produced gauge particle, the number density exhibits a stochastic resonance behavior.}
\label{fig-resonance-gauge}
\end{center}
\end{figure}

For the gauge mode the equation of motion is given by \eqref{eq:GaugeModesEOM}. Introducing the new variable
\begin{equation}\label{Ascript}
\mathcal{A}_k=a^{1/2} A_k,
\end{equation}
the equation takes the following form
\begin{equation}\label{A-eom}
\mathcal{A}_k^{\prime\prime}+{\tilde{\Omega}_{k}}^2 \mathcal{A}_k=0,
\end{equation}
where
\begin{equation}\label{Upsilonj}
\tilde{\Omega}_k^2\equiv \frac{k^2}{m^2 a^2}+\frac{\phi^2}{2\mu^2}(j^2+j)+\frac{1}{4}{\mathcal{H}}^2+1-\frac{a^{\prime\prime}}{2a}.
\end{equation}
Again  \eqref{A-eom} can be solved numerically imposing the Bunch-Davies vacuum in infinite past for the $\mathcal{A}_k$.

We have numerically solved \eqref{A-eom} for $k=0$. As it can be seen in Fig. \ref{fig-resonance-gauge} the gauge mode number density of produced particles, which is  given by \cite{preheat}
\begin{equation}\label{nA}
n_k^{\mathcal{A}}=\frac{1}{a^2}\left[\frac{\tilde{\Omega}_k}{2}\left(m^2 \frac{|\mathcal{A}_k^{\prime}|^2}{\tilde{\Omega}_{k}^2}+|\mathcal{A}_k|^2\right)-\frac{1}{2}\right].
\end{equation}
also demonstrates  stochastic resonance behaviour. Note that the $1/a^2$ factor in $n_k^{\mathcal{A}}$  will in principle cause the gauge mode particles to dilute.  The comoving number density of the particles overall increases more slowly due to the expansion of the universe. The production of gauge modes happens in the brane-antibrane inflation too \cite{Mazumdar:2008up}.

\section{Gravity Waves from preheating around the SUSY vacuum\label{sec:Gravitational-Waves-from}}

Effective preheating can lead to explosive particle creation and,
consequently, the production of stochastic Gravitational Waves (GWs) \cite{Khlebnikov:1997di}. The
latter arise from the tensor modes $h_{ij}$ of perturbations to the
FRW metric, and are linked to the former via the perturbed Einstein
equations,
\begin{equation}\label{eq:hijEOM}
\ddot{h}_{ij}+3H\dot{h}_{ij}-\left[\frac{\nabla^{2}}{a}+2\left(H^{2}+2\frac{\ddot{a}}{a}\right)\right]h_{ij}=\frac{16\pi G}{a^{2}}\delta S_{ij}^{\mathrm{TT}},
\end{equation}
where $\delta S_{ij}^{\mathrm{TT}}$ is the transverse-traceless  part of the
stress tensor perturbation $\delta S_{ij}=\delta T_{ij}-\frac{1}{3}\delta_{ij} \delta T_{k}{}^{k}$  which depends by construction on the number density and energy of the preheat fields. This stress-tensor perturbations are receiving contribution from the particles produced during the preheating era discussed in the previous section, which in turn source the gravity waves through \eqref{eq:hijEOM}.

Recalling that the Landau-Lifshitz pseudotensor \cite{Landau-Fields}
 associated with gravitational
radiation  is $T_{\mu\nu}=\langle h_{ij,\mu}h^{ij}{}_{,\nu}\rangle/32\pi G$,
we can write the ratio between the spectral energy density thereof
and the present-day total energy density as
\begin{equation}\label{eq:OmegaDefinition}
\Omega_{\mathrm{gw}}\left(f\right)=\frac{1}{\rho_{c}}\frac{\mathrm{d}}{\mathrm{d}\ln f}T_{00}=\frac{1}{\rho_{c}}\frac{\mathrm{d}}{\mathrm{d}\ln f}\sum_{i,j}\frac{1}{32\pi G}\left\langle h_{ij,0}^{2}\right\rangle ,
\end{equation}
where $f$ denotes the GW frequency. Using this, it is in principle
possible to compute the power spectrum, $\Omega_{\mathrm{gw}}h^{2}$.

Of course, the dynamics involved are highly nonlinear and  far
too complicated to render this task analytically tractable; instead,
we  resort to numerics. Thus, to determine the power spectrum
of GW generated during preheating after inflation by the various scalar
and gauge modes described in the previous section, we employ the lattice
simulator HLattice 2.0 \cite{Zhiqi}.

HLattice is generically designed to solve equations
of motion via a numerical scheme known as \emph{symplectic integration},
which is typically very stable and often used for long-term many-body
simulations in astronomy and particle physics. The basic idea of how
it works is as follows (for a detailed overview, the reader is referred
to \cite{Zhiqi}). Spatial
coordinates are discretized on a three-dimensional lattice -- in our cases, with
64 grid points along each edge-- and time evolution
is achieved by considering the Hamiltonian $\mathcal{H}$ of the system
which, in lieu of a spatial integral, can be written as a sum over
all of the lattice points. Then, any arbitrary function $F$ evolves
via
$$
\frac{\mathrm{d}F}{\mathrm{d}t}=\left\{ F,\mathcal{H}\right\} \equiv\mathbf{\hat{H}}F,
$$
where $\left\{ \cdot,\cdot\right\} $ is the Poisson bracket and $\mathbf{\hat{H}}$
is the corresponding functional operator. The solution is thus
\[
F\left(t+\mathrm{d}t\right)=\mathrm{e}^{\mathbf{\hat{H}}\mathrm{d}t}F(t).
\]
An $n$-th order symplectic integrator is constructed by factorizing
$\exp(\mathbf{\hat{H}}\mathrm{d}t)$ into a product of exponentials
of the constituent (kinetic and potential) terms of the Hamiltonian
up to $\mathcal{O}(\mathrm{d}t^{n+1})$. While HLattice 2.0 is in
principle able to implement this up to sixth order (using a fourth
order Runge-Kutta subintegrator, with a time step much smaller than
$\mathrm{d}t$, to solve the resulting equations of motion), we simply
used its second order symplectic integrator in obtaining all of the
results that follow, for the sake of keeping computational times manageable.

\subsection{GW from scalar modes}

\begin{figure}
\centering{\includegraphics[scale=0.75]{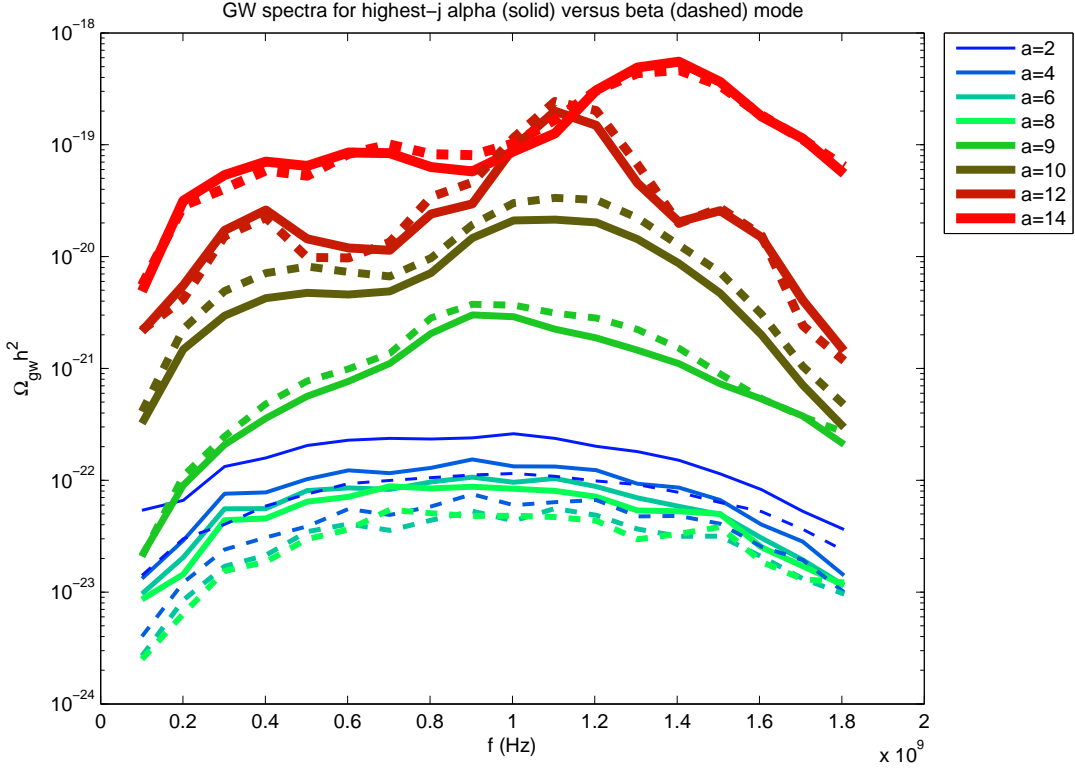}}
\caption{GW amplitude as a function of frequency due to the highest-$j$ scalar
modes, both $\alpha$ (solid) and $\beta$ (dashed), for a range of
scale factors from $a=1$ (beginning of preheating) to $a=14$.
\label{alpha-beta-single-mode}}
\end{figure}

\begin{figure}
\centering{\includegraphics[scale=0.60]{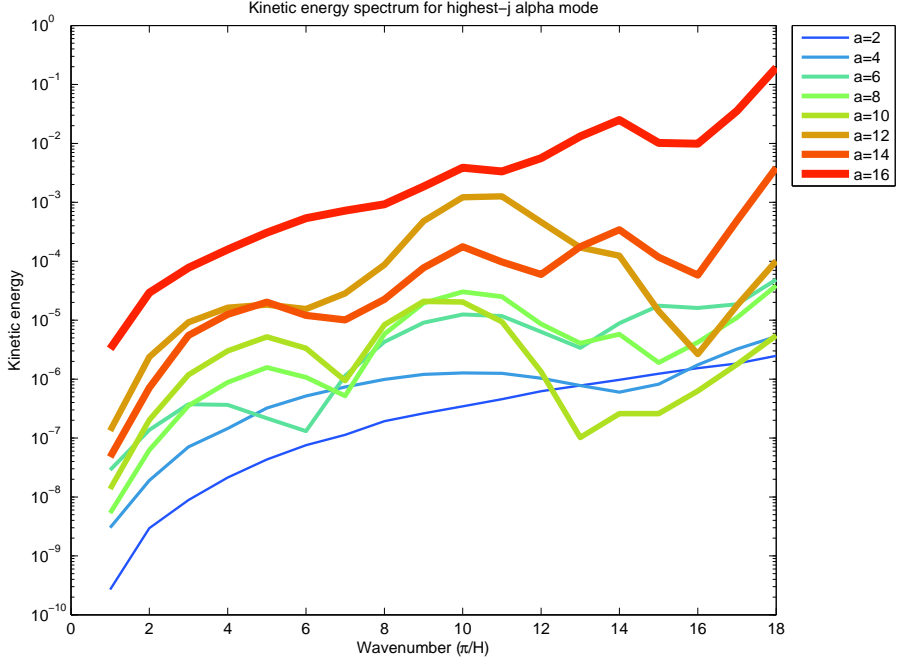}}
\vspace{-2mm}
\caption{{Kinetic spectrum of the highest $j$ $\alpha$ mode (in units of the background energy density) vs. the wavenumber in units of $\frac{\pi}{H}$. That is, the plot shows, $\frac{k^3}{4\pi^2}\left[\frac{k^2}{a^2} |\chi_k|^2\right]/(\rho_{\rm background})$ vs $\frac{k H}{\pi}$.}\label{kinetic-alpha}}
\end{figure}

The power spectra of GW due to the most massive--- {\it i.e.} highest $j$---scalar
modes (both $\alpha$ and $\beta$) are shown in Figure  \ref{alpha-beta-single-mode}.
The scale factor is normalized to $a=1$ at the end of inflation/beginning
of preheating, and we carry out the computation up to $a=14$, when the spectrum becomes UV dominated. Indeed, after preheating, field energies typically cascade
towards the UV,\footnote{Note that all simulations start out (small $a$) ``UV dominated'' and have larger energies at larger wave lengths. However, they do not remain so. But, there at a later time (larger $a$) which become UV dominated again.} and in HLattice this renders all further (higher $a$)
computations non-physical because of the finite resolution of the
simulator as well as its lacking treatment of quantum effects at very
high wavenumbers \cite{private}. To
illustrate this, we plot the kinetic energy spectrum of the highest
$j$ $\alpha$ mode in Fig. \ref{kinetic-alpha}
and observe that it starts to be dominated at the UV end for $a\geq14$.

We remark that, as is seen in Fig. \ref{alpha-beta-single-mode}, the two $\alpha$ and $\beta$ types of scalar preheat fields produce very
similar GW spectra, as may well be expected from inspecting their
masses (\ref{eq:MassSpectrumScalar}) and quartic couplings (\ref{eq:QuarticCouplings}):
For large $j$, both  $\alpha$ and $\beta$ type preheat fields have an approximate mass of
\begin{equation}
M_{\boldsymbol{\Psi}}^{2}\left(\phi\right)\approx\frac{1}{2}\lambda_{\mathrm{eff}}\phi^{2}j^{2},\label{eq:MApprox}
\end{equation}
and quartic coupling of
\begin{equation}
\Lambda_{\boldsymbol{\Psi}}\approx j^{2}\left[\frac{\lambda_{\mathrm{eff}}}{4}N\left(N^{2}-1\right)\right]\sum_{c=0}^{2j}\left(2c+1\right)\Big(\begin{array}{ccc}
j & j & c\\
1 & -1 & 0
\end{array}\Big)^{2}\Big\{\begin{array}{ccc}
j & j & c\\
\tfrac{N-1}{2} & \tfrac{N-1}{2} & \tfrac{N-1}{2}
\end{array}\Big\}^{2}.\label{eq:LambdaApprox}
\end{equation}
In producing these graphs we have assumed that $N=48000$. We have also taken the largest $j$ $\alpha$ and $\beta$ modes individually, {i.e.} $j=48000$ single $\beta$ and $\alpha$ mode.

\subsection{GW from gauge modes}

\begin{figure}
\centering{\includegraphics[scale=0.75]{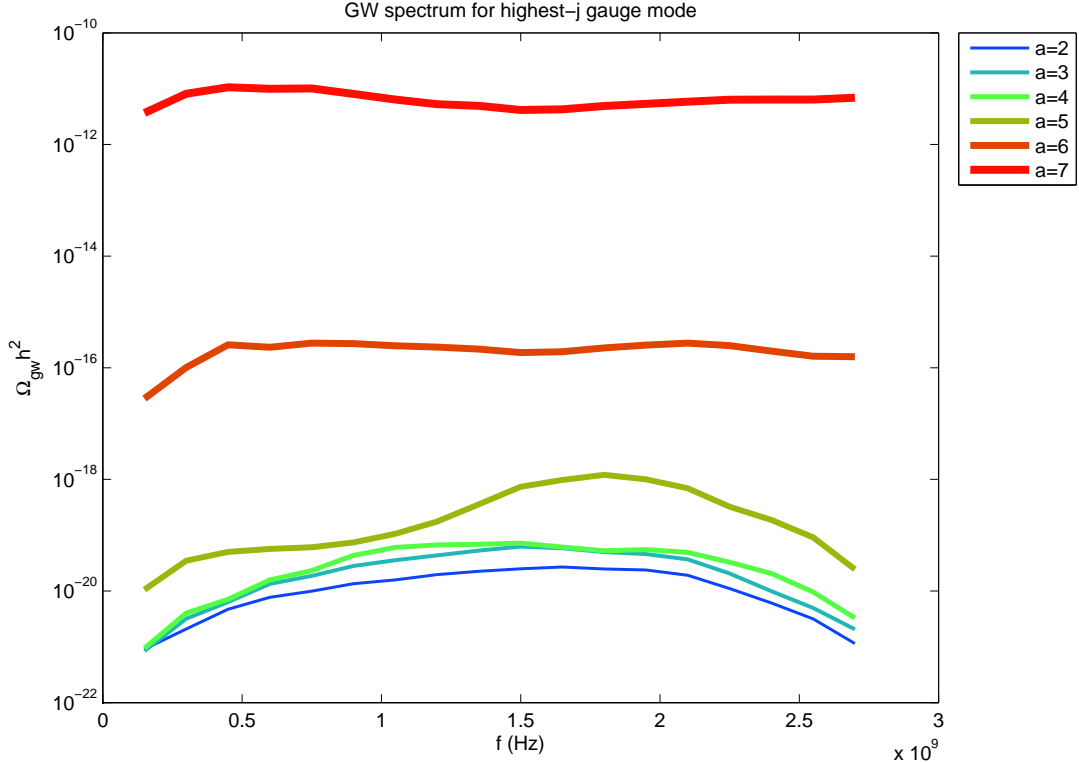}}
\vspace{-2mm}
\caption{{GW amplitude as a function of frequency due to the highest-$j$ gauge
mode, for a range of scale factors from $a=1$ (beginning of preheating)
to $a=7$.}\label{single-gauge-GW}}
\end{figure}

\begin{figure}
\centering{\includegraphics[scale=0.75]{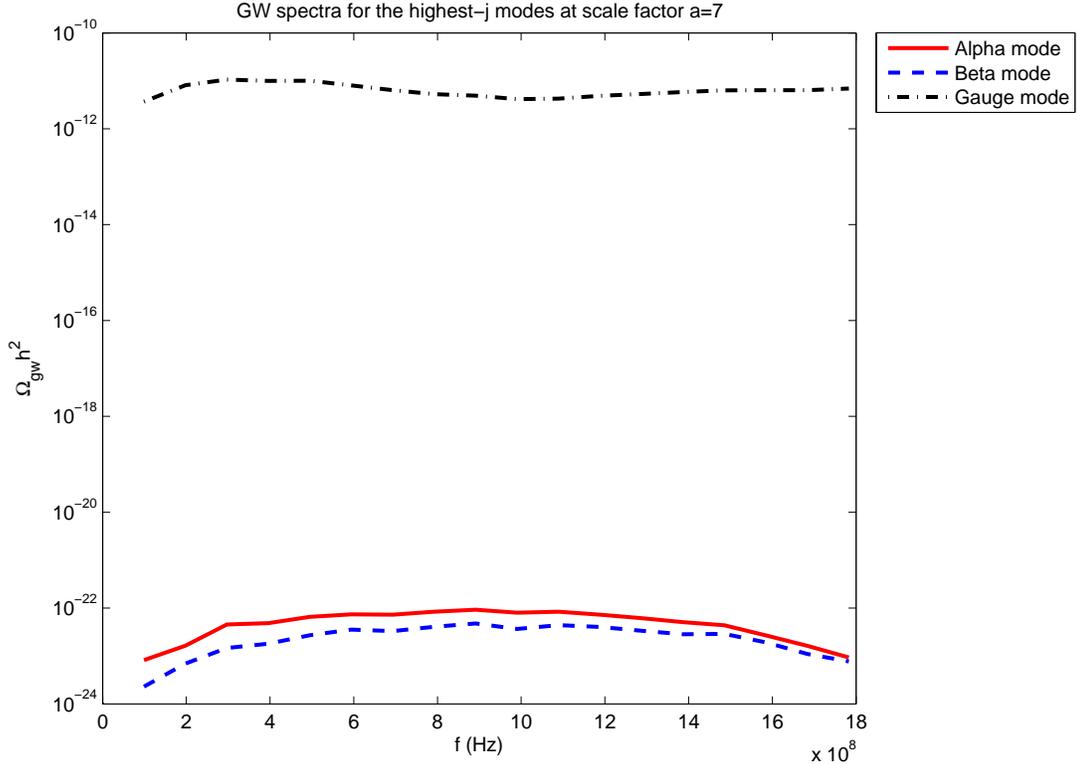}}
\vspace{-2mm}
\caption{{GW amplitude as a function of frequency due to the highest-$j$ modes,
both scalar ($\alpha$ in solid and $\beta$ in dashed) and gauge
(dashed-dotted), at scale factor $a=7$.}\label{gauge-vs-scalar}}
\end{figure}

The GW power spectrum due to the most massive gauge mode, up to $a=7$, before the UV domination kicks in,
is shown in Figure  \ref{single-gauge-GW}. Again we have focused on the largest $j$ gauge mode, $j=47999$. As in  the scalar mode case, the amplitude grows with increasing scale factor
under the clear effect of parametric resonance. However the growth
is much faster: amplitudes become as large as $10^{-11}$ by $a=7$, at which point the computations become UV dominated. The difference between gauge and scalar modes is essentially coming from the difference in their corresponding equations, and in particular the difference between $\Omega_k$ \eqref{omegal} and $\tilde\Omega_k$ \eqref{Upsilonj}. The delay in the enhancement of GW spectrum from scalar modes could be traced back to the fact that the presence of cubic coupling term in their equations of  motion generically slows down the resonance.
To compare the contributions to the total GW spectrum from the scalar
and gauge modes, they are plotted together in Figure \ref{gauge-vs-scalar}. The spectrum from a single gauge mode is also flatter in comparison with its scalar counterpart, but still a double hump feature of the gravity profile from preheating can be distinguished.

Thus, we see that the spectrum of GW produced by preheating following
M-flation is dominated by the gauge preheat fields, which give rise
to GW amplitudes more than 10 orders of magnitude greater (at a=7) than those
due to (either type of) their scalar counterpart.

\subsection{GW from several gauge modes}

As noted above the spectrum of GWs from the gauge modes dominate the scalar modes by a factor of $10$ orders of magnitude. This suggests that if all three modes are run together as the preheat fields, the gauge modes are more effective in the production of GWs.
However   this was done for a single scalar or gauge mode and at large $j$ there are $\sim 2j$ (for scalars) and $\sim 6j$ (for vectors) such modes for a given $j$. In principle one should consider the effects of all the degenerate modes. It may seem from \eqref{eq:hijEOM} and \eqref{eq:OmegaDefinition} that the GW power spectrum should grow like $j^2\sim N^2$. However,  given the highly nonlinear character of these equations this expectation can only hold for a very short time in the very low frequency region where the nonlinear effects are negligible. The larger the degeneracy, the earlier the UV domination, and hence modes have a shorter growth time. This is compatible with the analysis of \cite{Giblin:2010sp}. However one should note that in the study of \cite{Giblin:2010sp} the preheat modes are scalar fields, whereas the ones in our simulations are gauge modes, {\it i.e.} they appear with the friction term proportional to $H$, instead of $3H$ in the equations of motion.

Given the fact that for large $j$ gauge modes have a $6j$ degeneracy, to check the effects of degeneracies in our setup we should simulate the effect of $3\times95999=287997$ gauge mode as preheat fields. This number is quite huge and cannot be handled without substantial computational resources. To get an idea of the effects of degeneracy, we  tried three and six gauge modes\footnote{We should note that the simulation of a single mode with highest $j$-number up to the onset of UV domination took a week to perform on the Sharcnet cluster of the University of Waterloo. In comparison with previous studies on gravitational wave production from preheating, this is due to the large value of coupling of the inflaton to the preheat field which is of order $\lambda_{\rm eff}N^2\sim 1.6\times 10^{-4}$.}.

\begin{figure}[t]
\includegraphics[angle=0,
scale=0.75]{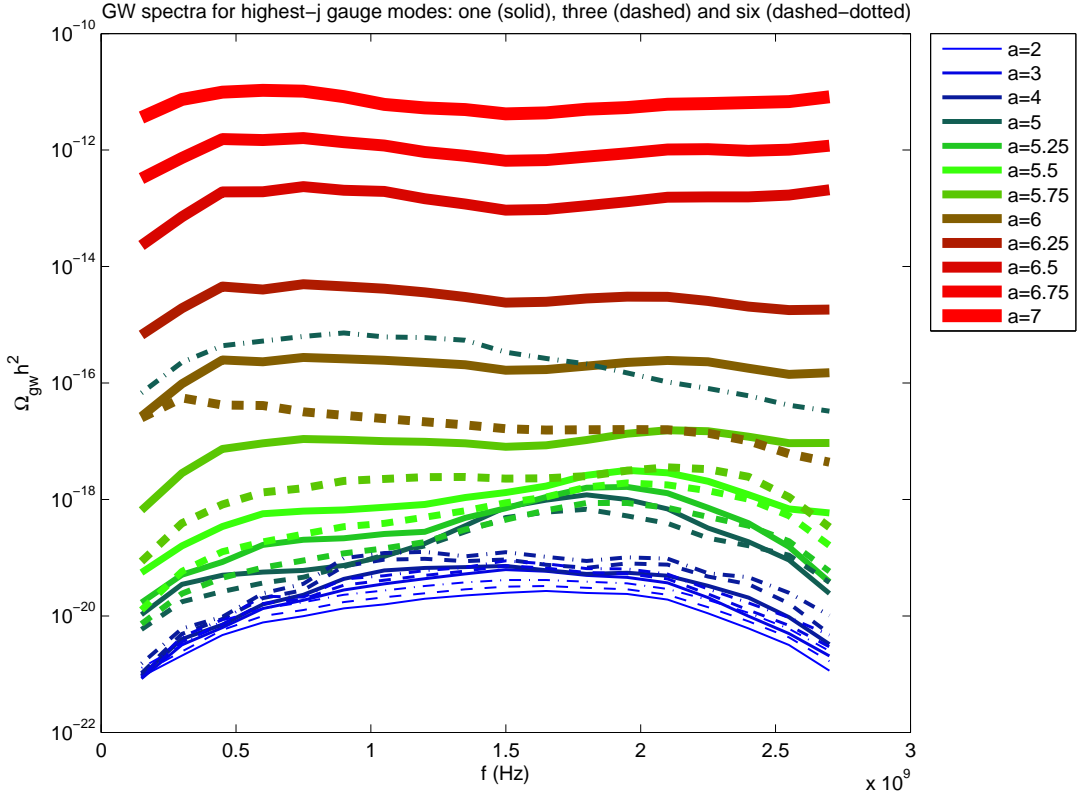}
\caption{Gravitational wave profile from one, three and six highest $j$ gauge modes until their corresponding onset of UV domination. The larger the numbers of preheat fields, the earlier the  the onset of UV domination.The spectrum generated from three and six gauge modes become steeper at the high frequency tail in comparison with the one of single mode.}
\label{six-gauge-modes}
\end{figure}

To explore the degeneracy effects more clearly we have shown the spectrum of GWs from one, three and six largest $j$ gauge modes in the same plot, Fig. \ref{six-gauge-modes}. Although these data are not enough for making a very sharp deduction, they still exhibit the following features:
\begin{itemize}
\item \textbf{Time dependence.} At the beginning of preheating, low $a$ up to $a=3$, the amplitude of the GW spectrum resulting from the three and six gauge preheat modes, is larger than that of single mode. As pointed out in \cite{Giblin:2010sp}, this is the stage the inflaton is coherently oscillating around its minimum and non-linear effects have not kicked in yet. However, as the inhomogeneities of the inflaton grow,  gravitational radiation is counteracted by the backreaction and the model with multiple preheat fields stops being efficient; nonlinear effects suppress the degeneracy effects and we see no large degeneracy effect. Moreover, UV domination happens earlier (at lower $a$) for larger degeneracy such that the amplitude  of GWs is almost degeneracy independent.
\item\textbf{Frequency dependence.}
Besides the amplitude of the produced GWs, frequency is the distinctive observational feature  in our model. Our current data with six gauge preheat modes already shows that the GWs of our model are in the $1-3$ GHz band and they are almost flat with amplitudes around $10^{-16}$.  Revealing the exact amplitude of the GW spectrum and its finer features in this range needs an analysis with a larger number of modes.
\end{itemize}

\section{Concluding remarks\label{sec:Conclusion}}

In this work we extended the analysis of \cite{M-flation,Ashoorioon:2011aa,M-flation-Landscape} on the M-flation model. As discussed, M-flation helps with the resolution of many of the principal theoretical difficulties
endemic to standard scalar field inflationary models. Moreover, M-flation is
also able to furnish concrete observational predictions courtesy of
its built-in preheating mechanism around the $\phi=0$ vacuum. In search for possible, beyond CMB, observational signatures of M-flation we have analyzed
the power spectra of gravitational waves produced in this model due to the different types
of its preheat fields. We have found
that the gauge preheat fields contribute overwhelmingly to this process as compared
to their scalar counterparts, producing a large amplitude spectrum in the few GHz
band with an amplitude of order $10^{-16}$. It is hoped that such a spectrum could be observed by ultra-high
frequency GW detectors that may be able to probe the GHz band, such
as the Birmingham HFGW resonant antenna \cite{Birmingham} or the one at Chongqin University \cite{Chongqin}. The Birmingham detector works based on the detection of the rotation of the polarization vector of an electromagnetic wave induced
by the interaction between a gravitational wave and the polarization vector of the electromagnetic
wave. The sensitive frequency range is at $10^8$ HZ. The Chongqing detector exploits the electromagnetic interaction of a Gaussian beam
propagating through a static magnetic field. These detectors work based on different
principles from the phase measurement with the laser interferometry developed in
the ground-based large-scale interferometers around few hundred Hz.

One should note that the GW spectrum we discussed in this paper is in the high frequency range, and is in addition to the spectrum of gravity waves (tensor modes) that the model produces at the CMB scales, with the tensor-to-scalar ratio $r\simeq 0.048$ \cite{M-flation}. In addition the lightest spectator mode in this inflationary region will create a substantial amplitude of isocurvature perturbations with amplitude $P_{\mathcal{S}}/P_{\mathcal{R}}\simeq 5\times 10^{-3}$ which has a degeneracy of three \cite{M-flation}.\footnote{We note that the \emph{Planck} bound is $P_{\mathcal{S}}/P_{\mathcal{R}}<3.6\times 10^{-2}$ \cite{Planck}.} These features could be used to distinguish M-flation in this region from other inflationary models.

\section*{Acknowledgements}

We are greatly indebted to Z. Huang for his extensive help and advice
in our implementation of the HLattice 2.0 code used here. This work
was supported in part by the Natural Sciences and Engineering Research
Council of Canada. M. O. acknowledges additional support from the
University of Waterloo Department of Physics and Faculty of Mathematics. A. A. is supported by the
Lancaster-Manchester-Sheffield Consortium for Fundamental Physics under STFC grant ST/J000418/. A. A. also acknowledges the hospitality of Uppsala Institute for Theoretical Physics during the completion of this work.

\bibliographystyle{apsrev}

\end{document}